\documentclass[12pt]{JHEP3}
\usepackage{graphicx}
\usepackage{dcolumn}  
\usepackage{bm}    
\usepackage{amssymb} 
\usepackage{amsmath,bm}
\usepackage{amsfonts}    
\usepackage{slashed}  
\usepackage[mathscr]{euscript}
\title{Higher spin fermions in the BTZ black hole}
\author{Shouvik Datta,  Justin R. David   \\
 Centre for High Energy Physics,
Indian Institute of Science,\\ C.V. Raman Avenue, Bangalore 560012, India. \\
\email{shouvik, justin@cts.iisc.ernet.in}\\
}

\abstract{ 
Recently it has been shown that 
the wave equations of  bosonic higher spin fields in the BTZ background 
can be solved exactly. In this work we extend this analysis to fermionic higher spin fields. 
We solve the wave equations for arbitrary half-integer spin fields in the BTZ black hole background and obtain exact expressions for their quasinormal modes. These quasinormal modes are shown to agree  precisely with the poles of the corresponding two point function in the dual conformal field theory as predicted by the AdS/CFT correspondence. We also obtain an expression for the 1-loop determinant in terms of the quasinormal modes and show it agrees with that obtained by integrating  the heat kernel found by 
group theoretic methods.
}

\begin{document}
\def\G{\Gamma}
\def\lb{\left(}
\def\rb{\right)}
\section{Introduction}

Studying the behaviour of various fields in black hole backgrounds have always 
revealed useful information of the nature of black  holes.  Solving the equation of 
motion of various fields  in the black hole background 
is usually the first step in understanding properties like Hawking radiation and
 quasinormal modes of the black hole. 
In most  situations  the equations
of motion  are solvable only numerically or in some limits. 
The BTZ black hole is a rare example of  a black hole background 
in which  the equations of motion of all 
integer spins are  solvable in closed form \cite{Datta:2011za}. 
in terms of hypergeometric functions.  
In \cite{Datta:2011za} the complete spectrum of quasinormal modes of all integer spin fields 
in the BTZ background  was obtained and they were shown to agree with the prediction from 
$AdS_3/CFT_2$. 
The aim of this paper is to generalize this result to the case of 
arbitrary half integer spins.
Thus the BTZ black hole is perhaps  the only example of a black hole in which
 spectrum of excitations of arbitrary spin fields 
can be completely solved in closed form.  In fact classical string propagation in 
the BTZ background is shown to be integrable \cite{David:2011iy}. 
All these observations suggest that it is possible to obtain the complete spectrum of 
string excitations. 

Arbitrary massive half integer spins in 3 dimensional backgrounds 
 are present  in all  examples of  $AdS_3/CFT_2$ that occur in 
string theory. A prototype example of this is the D1-D5 system and the corresponding BTZ black hole
\cite{Maldacena:1998bw}. 
Arbitrary massless half integer higher spins also occur in the supersymmetric
versions  of the $AdS_3/CFT_2$  duality proposed by 
Gaberdiel and Gopakumar \cite{Gaberdiel:2010pz} which has recently been studied in \cite{Creutzig:2011fe}. 
Let us briefly recall the  $AdS_3/CFT_2$ dictionary regarding a field of spin $s$. 
$AdS_3/CFT_2$ duality relates a spin $s$ field propagating 
in $AdS_3$ to an operator ${\cal O}$ in the dual conformal field theory characterized by 
conformal weights $(h_L, h_R)$ with \cite{Aharony:1999ti}
\begin{equation}
 h_R- h_L = \pm s.
\end{equation}
The mass of the propagating field $m$ is related to the conformal dimension of the 
operator ${\cal O}$ which is given by 
\begin{equation}
h_R + h_L = \hat \Delta.
\end{equation}
When the  conformal field theory is at finite temperature it  is dual to the BTZ black hole. 
Then the poles 
of  the retarded two point function of the operator ${\cal O }$ are given by 
the quasinormal modes  of the spin $s$ field \cite{Birmingham:2001pj,Son:2002sd,Kovtun:2005ev}. 
The two point function of the operator ${\cal O }$ is determined  
by conformal invariance. The poles of the  retarded Green's
function in the complex frequency plane are then given by 
\cite{Gubser:1997cm,Birmingham:2001pj}
\begin{equation}
\label{locp}
 \omega_L = k -4\pi i T_L( \hat n + h_L), \qquad \omega_R = -k -4\pi i T_R(\hat n  + h_R). 
\end{equation}
$\omega$ and $k$ refer to the frequency and momentum respectively and 
$\hat n = 0 , 1, 2, \ldots$. 
The $L_, R$ subscripts denote left and right moving poles  $T_L, T_R$ are the left and 
right moving temperatures of the CFT which are related to the temperature and the angular
potential of the BTZ black hole.  
In \cite{Datta:2011za} we have verified this prediction 
for arbitrary integer spins by explicitly solving the wave equations of the higher 
spin field and obtaining their quasi-normal modes. 
In this paper we solve the wave equations of arbitrary half integer spins
in the BTZ background and obtain their quasinormal modes. It will be shown 
that the quasinormal modes coincide with the location of the poles predicted by the 
$AdS_3/CFT_2$ correspondence given in (\ref{locp}). 
Thus this work completes the analysis begun in \cite{Datta:2011za}. 
It is indeed remarkable that the wave equations of arbitrary spins
can be solved in closed form for the BTZ background. 

An interesting property of quasi-normal modes discovered recently,  is that the 1-loop 
determinant of the corresponding field can be obtained  by considering suitable products of the 
quasi-normal modes \cite{Denef:2009kn}. We show that  by considering suitable products of the 
quasi-normal modes of the half integer spin fields we reproduce the 
one loop determinant constructed by integrating the heat kernel in thermal $AdS_3$ found
by group theoretic methods in \cite{David:2009xg}.

The organization of this paper is as follows. We will first review the description of 
higher spin fermions in $AdS_3$.  This will enable us to introduce the notations
and the conventions we use in this paper. 
In section 3 we will show that the equations of motion of  arbitrary half integer  spin fields in the 
BTZ background can be simplified and solutions can be found in closed from 
in terms of hypergeometric functions. This generalizes the work of 
\cite{Das:1999pt} which solves the wave equations for the $s=1/2$ case. 
Once the solutions have been found we extract their quasi-normal modes and 
show that it agrees with that given in (\ref{locp}). 
In section 4 we write down the one-loop determinant for the higher spin fermions in 
terms of products over the corresponding quasi-normal modes and show that it agrees
with that evaluated by group theoretic methods. 
Appendix A contains some details about the geometry of the BTZ black hole 
and the properties of Laplacians of higher spin fermions. 
Appendix B contains the proof an identity which is required in our analysis of the 
wave equations for the higher spin fermions.

\def\nn{\nonumber}
\def\kp{k_+}
\def\km{k_-}
\section{Description of higher spin fermions in $AdS_3$}

The massive higher spin fermion fields with $s=n+\frac{1}{2}$ are realized by 
completely symmetric tensors of rank $n$, $ \Psi _{\mu_1\mu_2\dots\mu_n} $.
They satisfy the following equations \cite{Buchbinder:2007vq,Metsaev:2003cu}.
\begin{eqnarray} \label{hsf1}
[\Gamma ^\mu \nabla _\mu - m_n] \Psi _{\mu_1\mu_2\dots\mu_n} &=&0, \\ 
\label{gcond}
\nabla^\mu \Phi_{\mu \mu_2\cdots \mu_n} &=& 0,  \\
\label{tr}
\Gamma ^\mu \Psi _{\mu\mu_2\dots\mu_n} &=&0. 
\end{eqnarray}
Here $\Psi_{\mu_1, \mu_2\cdots \mu_n}$ is a two component Dirac fermion which
is totally symmetric in the indices $\mu_1, \mu_2\cdots \mu_s$.  
The mass of the spin \( (n+\frac{1}{2})\) field in  $AdS_3$  is denoted as $m_n$. 
It is usually written as the following sum  \cite{Buchbinder:2007vq}
\begin{equation}
m_n = \left(n-\frac{1}{2}\right) + M. 
\end{equation}
where the first term is due to the curvature of $AdS_3$ and $M$ is the actual mass of the 
field.  Here we have set the radius of $AdS_3$ to unity. 
The curved space Dirac matrices obey the Clifford algebra.
\begin{equation}\label{cliff}
\{\Gamma_\mu,\Gamma_\nu\}= 2 g_{\mu\nu}. 
\end{equation}
Here the $\Gamma ^\mu $ is written using   the vierbien ($e_a^\mu$) by $\Gamma ^\mu = \gamma ^a e _a^\mu$. We use the mostly $+$ convention for the metric 
$g_{\mu, \nu}$ and $\mu, \nu, a  \in\{0, 1, 2\}$.
The covariant derivative is defined as 
\begin{eqnarray} \label{defcova}
 \nabla_\mu \Psi _{\mu_1\mu_2\cdots\mu_n} &=& 
\partial_\mu \Psi _{\mu_1\mu_2\cdots\mu_n}+ 
\frac{1}{8} \omega_{\mu}^{ab}[\gamma_a, \gamma_b] \Psi _{\mu_1\mu_2\cdots\mu_n}\\ \nonumber 
& & -
\tilde\Gamma ^\rho_{\mu\mu_1} \Psi_{\rho, \mu_2\cdots \mu_n}- \cdots  
- \tilde \Gamma^\rho_{\mu\mu_n} \Psi_{\mu_1 \cdots \rho}, 
\end{eqnarray}
where  $\tilde\Gamma$ are the Christoffel symbols and $\omega_{\mu}^{ab}$ refer to the 
spin connection. 
 Equation (\ref{tr}) and (\ref{cliff}) also imply 
\begin{equation}\label{double-trace}
g^{\mu \nu}\Psi _{\mu\mu_2\dots\mu_n} =0. 
\end{equation}

One  important fact of higher spin fermionic fields 
in $AdS_3$ which will be important for our 
analysis is that the Dirac equation (\ref{hsf1}) and the 
gauge condition (\ref{gcond})  is equivalent  to the 
following first order equation \cite{Tyutin:1997yn} together with the traceless condition
(\ref{tr}) for $m \neq 0$. 
\begin{equation}\label{hs-cs}
\Gamma^{\mu\nu\rho} \nabla _\nu \Psi _{\rho\mu_2\dots\mu_n} + m \Gamma ^{\mu\nu}  \Psi _{\nu\mu_2\dots\mu_n} =0, 
\end{equation}
where
\begin{eqnarray}
\label{g1}
\Gamma^{\mu\nu} &= & \frac{1}{2} (\Gamma^\mu\Gamma^\nu-\Gamma^\nu\Gamma^\mu),   \\
\nonumber
  &=&  \Gamma^\mu\Gamma^\nu-g^{\mu\nu}. 
\end{eqnarray}
The second line of the above equation is obtained by using the relation in (\ref{cliff}). 
Similarly 
\begin{eqnarray}
\label{g2}
\Gamma^{\mu\nu\rho} &=&  \frac{1}{3!} (\Gamma^\mu\Gamma^\nu\Gamma^\rho - \Gamma^\nu\Gamma^\mu\Gamma^\rho + \text{et.  cycl.}),  \\ \nonumber
   &=& 
 \Gamma^\mu\Gamma^\nu\Gamma^\rho - g^{\mu\nu}\Gamma^\rho -
 g^{\nu\rho}\Gamma^\mu + g^{\mu\rho}\Gamma^\nu. 
\end{eqnarray}
where again we have repeatedly used (\ref{cliff}) to obtain the second line.  Note that 
for $n=1$ the equation in (\ref{hs-cs}) reduces to the Rarita-Schwinger equation 
for the gravitino. 
We shall now show that the covariant gauge condition 
(\ref{gcond}) 
is automatically implied once we have (\ref{hs-cs}) and (\ref{tr}). Using (\ref{g2}) in (\ref{hs-cs}) and contracting with $\Gamma_\mu$ we obtain
\begin{align}
\Gamma_\mu(\Gamma^\mu\Gamma^\nu\Gamma^\rho - g^{\mu\nu}\Gamma^\rho - g^{\nu\rho}\Gamma^\mu + g^{\mu\rho}\Gamma^\nu) \nabla_\nu \Psi_{\rho\mu_2\dots\mu_n} + m\Gamma_\mu (\Gamma^\mu\Gamma^\nu-g^{\mu\nu})  \Psi_{\nu\mu_2\dots\mu_n} =0. 
\end{align}
Using the fact $\Gamma_\mu\Gamma^\mu = 3$ we get 
\begin{equation}
[\slashed{\nabla} (\Gamma^\rho  \Psi_{\rho\mu_2\dots\mu_n} ) + \nabla^\rho  \Psi_{\rho\mu_2\dots\mu_n}] + 2m \Gamma^\nu  \Psi_{\nu\mu_2\dots\mu_n} =0 .
\end{equation}
Imposing the tracelessness condition (\ref{tr}) results in 
\begin{equation} \label{cov-gauge}
 \nabla^\rho  \Psi_{\rho\mu_2\dots\mu_n} =0, 
\end{equation}
for $m\neq 0$. 
We will now  show that the equation (\ref{hs-cs}) is equivalent to (\ref{hsf1}) when we have 
tracelessness (\ref{tr}) and the gauge condition (\ref{cov-gauge}). 
Using (\ref{g2}) in (\ref{hs-cs}) we obtain,
\begin{align} \label{sim}
(\Gamma^\mu\Gamma^\nu\Gamma^\rho - g^{\mu\nu}\Gamma^\rho - g^{\nu\rho}\Gamma^\mu + g^{\mu\rho}\Gamma^\nu) \nabla_\nu \Psi_{\rho\mu_2\dots\mu_n} + m(\Gamma^\mu\Gamma^\nu-g^{\mu\nu})  \Psi_{\nu\mu_2\dots\mu_n} =0.
\end{align}
In the first pair of parentheses, the first three terms do not contribute because of (\ref{tr}) and (\ref{cov-gauge}) respectively. Similarly the first term in the second pair of parentheses do not contribute due to (\ref{tr}).  Thus on multiplying by $g_{\mu\mu_1}$ we obtain
\begin{equation}\label{cs2}
\slashed{\nabla}\Psi_{\mu_1\mu_2\dots\mu_n} - m \Psi_{\mu_1\mu_2\dots\mu_n} =0. 
\end{equation}
This is clearly equivalent to (\ref{hsf1}) with the following relation between masses. 
\begin{equation}
m=m_n. 
\end{equation}
For definiteness we will take $m>0$ and it will be seen subsequently that this leads to the 
situation with $h_L- h_R = s$. The analysis can be carried out for the case with 
$m<0$ and it will lead to the situation with $h_L - h_R =-s$. 

Note that using (\ref{g1})  and (\ref{tr}) 
the Chern-Simons like equation in (\ref{hs-cs}) can also be written as
\begin{equation}
\label{hs-cs2}
g_{\mu\sigma} \Gamma^{\mu\nu\rho}\nabla_\nu \Psi_{\rho\mu_2\dots\mu_n} 
-m \Psi_{\sigma\mu_2\dots\mu_n} =0 . 
\end{equation}
As a final consistency check we show that (\ref{hs-cs2}) agrees with the symmetry of the tensor. 
The above equation must be symmetric under the exchange of $\sigma$ and $\mu_2$. 
This equivalent to saying
\begin{equation}
\epsilon^{\sigma\mu_2\eta}g_{\mu\sigma} \Gamma^{\mu\nu\rho}\nabla_\nu \Psi_{\rho\mu_2\dots\mu_n} =0 , 
\end{equation}
 should be true. Let us examine the RHS of the above equation
\begin{eqnarray}
 \epsilon^{\sigma\mu_2\eta}g_{\mu\sigma} \Gamma^{\mu\nu\rho}\nabla_\nu \Psi_{\rho\mu_2\dots\mu_n} 
&=& \epsilon_\mu^{\ \mu_2\eta}  \Gamma^{\mu\nu\rho}\nabla_\nu \Psi_{\rho\mu_2\dots\mu_n} , \nn \\
&=& \epsilon_\mu^{\ \mu_2\eta} g^{\mu\rho} \slashed{\nabla} \Psi_{\rho\mu_2\dots\mu_n},  \nn  \\
&=& \epsilon^{\rho \mu_2\eta}  \slashed{\nabla} \Psi_{\rho\mu_2\dots\mu_n} . \nn \\
&=&  0 \nn 
\end{eqnarray}
In the second step we have used similar manipulations as those used in simplifying (\ref{sim}). 

The BTZ black hole is obtained by identifications of $AdS_3$ \cite{Banados:1992gq}. Thus it is locally 
$AdS_3$ and the above analysis for the equations of motion for the fermionic higher spin fields
in $AdS_3$  can be carried over to the BTZ background. 
We will use the following metric for the BTZ black hole
\begin{equation}
\label{dbtz}
ds^2 = d\xi^2 - \sinh^2 \xi dx_+^2 + \cosh^2\xi dx_-^2. 
\end{equation}
The horizon is at $\xi =0$ and the boundary is at $\xi =\infty$. 
The relation between these co-ordinates and the conventional co-ordinates is reviewed
in Appendix A. The appendix also lists other properties of this background which will 
be used in the paper. 
We will use the following convention for the flat space Dirac matrices \cite{Das:1999pt}
\begin{equation}
\label{defgamma}
\gamma^0 = i \sigma^2, \qquad \gamma^1  = \sigma^1, \qquad
\gamma^2 =  \sigma^3.
\end{equation}
where $\sigma$'s refer to the Pauli matrices
The vierbein for the BTZ metric in (\ref{dbtz}) is given by 
\begin{equation}\label{vier}
e_0   = - \sinh\xi dx_+, \qquad e_1 = d\xi, \qquad e_2 = \cosh\xi dx_-. 
\end{equation}
The spin connection is given by 
\begin{eqnarray}
\hat \omega_+ &=& \frac{1}{8} \omega_{+}^{ab}[\gamma_a, \gamma_b] =
- \frac{1}{2} \cosh\xi \sigma^3, \\ \nonumber
\hat \omega_- &=& \frac{1}{8} \omega_{+}^{ab}[\gamma_a, \gamma_b] = 
\frac{i }{2} \sinh \xi \sigma^2, \\ \nonumber
\omega_{\xi}^{ab} &=& 0. 
\end{eqnarray}

\section{Solving higher spin fermion equations}

In this section we solve the fermionic higher spin equations (\ref{hsf1}) 
in the background of the BTZ black hole and obtain their quasi-normal modes.
Our strategy will be the following:
Note that due to the traceless condition (\ref{tr}) we can restrict our attention 
to components of the higher spin fermion whose indices lie along 
the $+$ and $-$ directions. 
The tracelessness condition (\ref{tr}) for the BTZ metric in (\ref{dbtz})  is given by 
\begin{equation}
\gamma^0 \frac{1}{\sinh\xi}\Psi_{+\mu_2\dots\mu_n} + \gamma^1 \Psi_{\xi\mu_2\dots\mu_n} + \gamma^2 \frac{1}{\cosh\xi}\Psi_{-\mu_2\dots\mu_n}=0, 
\end{equation}
from which we obtain
\begin{equation}\label{tr2}
 \Psi_{\xi\mu_2\dots\mu_n} = 
-\ \gamma^1 \gamma^0 \frac{1}{\sinh\xi}\Psi_{+\mu_2\dots\mu_n}  - \gamma^1\gamma^2 \frac{1}{\cosh\xi}\Psi_{-\mu_2\dots\mu_n}. 
\end{equation}
From the above equation it is clear 
any component with whose indices lie along the radial coordinate $\xi$ can be expressed
in terms of components along $+$ and $-$ directions by the  repeated use of (\ref{tr2}). 
The next step in our analysis  which is carried out in 
section 3.1 consists of showing that the 
spin-$(n+\frac{1}{2})$ Dirac operator on components with indices only along $+$ and $-$ directions 
reduces to  that of the spin $\frac{1}{2}$ Dirac operator but with a mixing `mass  matrix'.
This property is similar to the observation seen for the bosonic higher spin fields in 
\cite{Datta:2011za} where the spin $s$ Laplacian  reduced to the scalar Laplacian with a 
mixing mass matrix.   In  section 3.2  the equations are diagonalized and solved in 
terms of hypergeometric functions up to a constant which depends on the 
polarization of the higher spin component. 
We then show that all the  polarizations can be  related to a single constant using
  (\ref{hs-cs2}). In section 3.3 we determine the quasi-normal modes
and show that they coincide with the poles of the corresponding
two point function as expected from the $AdS_3/CFT_2$ correspondence. 

\subsection{Reducing the  higher spin Dirac operator  to the spin 
$\frac{1}{2}$  Dirac operator }

The  action of 
the Dirac operator $\slashed{\nabla}$ on 
$\Psi_{\mu_1\mu_2\dots\mu_p\nu_1\nu_2\dots\nu_q}$ is given by 
\begin{align} \label{spind}
\slashed{\nabla} &\Psi_{\mu_1\mu_2\dots\mu_p\nu_1\nu_2\dots\nu_q} \nn \\
=& \Gamma ^\mu ( \partial_\mu + \omega_\mu )  \Psi_{\mu_1\mu_2\dots\mu_p\nu_1\nu_2\dots\nu_q} \nn \\
&- \Gamma^\mu (\tilde\Gamma^\eta_{\mu\mu_1}\Psi_{\eta\mu_2\dots\mu_p\nu_1\nu_2\dots\nu_q} 
+ \tilde\Gamma^\eta_{\mu\mu_2}\Psi_{\mu_1\eta\mu_3\dots\mu_p\nu_1\nu_2\dots\nu_q} + \cdots 
+ \tilde\Gamma^\eta_{\mu\mu_p}\Psi_{\mu_1\dots\mu_{p-1}\eta\nu_1\nu_2\dots\nu_q}  \nn \\
&\quad \qquad + \tilde\Gamma^\eta_{\mu\nu_1}\Psi_{\mu_1\mu_2\dots\mu_p\eta\nu_2\dots\nu_q}
 + \tilde\Gamma^\eta_{\mu\nu_2}\Psi_{\mu_1\mu_2\dots\mu_p\nu_1\eta\nu_3\dots\nu_q} + \cdots 
+\tilde\Gamma^\eta_{\mu\nu_q}\Psi_{\mu_1\dots\mu_p\nu_1\nu_2\dots\nu_{q-1}\eta} ).
\end{align}
As we have argued earlier it is sufficient to examine the components 
with indices along the $+$ and $-$ coordinates. 
Therefore we choose $\mu_1\cdots \mu_p = +\cdots +$ and $\nu_1\cdots \nu_q = -\cdots -$. 
We also introduce the notation `$(p)$' to mean $p$ number of $+$ indices and 
by we mean `$(q)$' number of $-$ indices with $p+q = n$. 
For example consider a spin-$\frac{11}{2}$  field with $n=5$ then
\begin{equation}
\Psi_{(2)(3)} = \Psi_{++---}.
\end{equation}
Note that 
the operator occurring in the first term  of (\ref{spind}) is same as  the 
Dirac operator $\slashed{\nabla}$ acting on the spin-$\frac{1}{2}$ object $\Psi$. We denote this as $\Delta \Psi_{ (p)(q)}$.  Let us define
\def\ma{\mathcal{A}} 
\begin{equation}
\begin{aligned}
\ma_{\mu (p)(q)} =  &\tilde\Gamma^\eta_{\mu\mu_1}\Psi_{\eta\mu_2\dots\mu_p\nu_1\nu_2\dots\nu_q} + \tilde\Gamma^\eta_{\mu\mu_2}\Psi_{\mu_1\eta\mu_3\dots\mu_p\nu_1\nu_2\dots\nu_q} + \cdots 
+ \tilde\Gamma^\eta_{\mu\mu_p}\Psi_{\mu_1\dots\mu_{p-1}\eta\nu_1\nu_2\dots\nu_q}   \\
 &+ \tilde\Gamma^\eta_{\mu\nu_1}\Psi_{\mu_1\mu_2\dots\mu_p\eta\nu_2\dots\nu_q} + 
 \tilde\Gamma^\eta_{\mu\nu_2}\Psi_{\mu_1\mu_2\dots\mu_p\nu_1\eta\nu_3\dots\nu_q} + \cdots 
+ \tilde\Gamma^\eta_{\mu\nu_q}\Psi_{\mu_1\dots\mu_p\nu_1\nu_2\dots\nu_{q-1}\eta}  \\
= &p\tilde\Gamma^\eta_{\mu +}\Psi_{\eta(p-1)(q)}+ q\tilde\Gamma^\eta_{\mu -}\Psi_{\eta(p)(q-1)}. 
\end{aligned}
\end{equation}
Thus with these notations and definitions we have
\begin{equation}\label{dirac-opr}
\slashed{\nabla}  \Psi_{ (p)(q)} = \Delta \Psi_{ (p)(q)} -  \Gamma^\mu \ma_{\mu(p)(q)}.
\end{equation}
where the first term is the ordinary Dirac operator and 
the second term of (\ref{dirac-opr}) is given by 
\begin{align}\label{ga}
 \Gamma^\mu \ma_{\mu(p)(q)} =& \gamma ^a e^\mu_a  \ma_{\mu(p)(q)},  \nn \\
 =&  \gamma ^0 e^{\ +}_0  \ma_{+(p)(q)} +  \gamma ^1 e^{\ \xi}_1  \ma_{\xi(p)(q)} +  \gamma ^2 e^{\ -}_2  \ma_{-(p)(q)},  \nn \\
 =&  \gamma ^1  \ma_{\xi(p)(q)} +  \gamma ^0 \frac{1}{\sinh\xi}  \ma_{+(p)(q)} +  \gamma ^2 \frac{1}{\cosh\xi}  \ma_{-(p)(q)}. 
\end{align}
We will now  evaluate each of the  components of $\ma$'s explicitly: 
We begin with the component $\ma_{\xi(p)(q)}$ which is given by 
\begin{align}\label{az}
 \ma_{\xi(p)(q)}  =\ & p\tilde\Gamma^+_{\xi +}\Psi_{+(p-1)(q)} + q 
 \tilde\Gamma^-_{\xi -}\Psi_{-(p)(q-1)},  \nn \\
 =\ & (p \coth\xi  + q \tanh\xi )\Psi_{(p)(q)}. 
 \end{align}
 Evaluating $\ma_{+(p)(q)}$ leads to 
\begin{align}\label{ap}
 \ma_{+(p)(q)}  =\ & p\tilde\Gamma^\eta_{+ +}\Psi_{\eta(p-1)(q)} + q 
 \tilde\Gamma^\eta_{+ -}\Psi_{\eta(p)(q-1)},  \nn \\
 =\ &  p\tilde\Gamma^\xi_{+ +}\Psi_{\xi(p-1)(q)} 
= p\ {\cosh\xi\sinh\xi }\Psi_{\xi(p-1)(q)}, \nn \\
=\ & p\ {\cosh\xi\sinh\xi }(  -\gamma^1 \gamma^0 \frac{1}{\sinh\xi}\Psi_{+(p-1)(q)}  - \gamma^1\gamma^2 \frac{1}{\cosh\xi}\Psi_{-(p-1)(q)}), 
\nn \\
=\ &  -p\gamma^1 \gamma^0 \cosh\xi\Psi_{(p)(q)}  -p \gamma^1\gamma^2 \sinh\xi\Psi_{(p-1)(q+1)}. 
 \end{align}
 To obtain the last line in the above equation we have used the condition 
 (\ref{tr2}). 
 Finally  $\ma_{-(p)(q)}$ is given by 
\begin{align}\label{am}
 \ma_{-(p)(q)}  &= 
 p\hat\Gamma^\eta_{- +}\Psi_{\eta(p-1)(q)} + q \tilde\Gamma^\eta_{- -}\Psi_{\eta(p)(q-1)},  \nn \\
 &=   q\tilde\Gamma^\xi_{--}\Psi_{\xi(p)(q-1)} 
= -q {\cosh\xi\sinh\xi }\Psi_{\xi(p)(q-1)},  \nn \\
&=  -q {\cosh\xi\sinh\xi }(  -\gamma^1 \gamma^0 \frac{1}{\sinh\xi}\Psi_{+(p)(q-1)}  - \gamma^1\gamma^2 \frac{1}{\cosh\xi}\Psi_{-(p)(q-1)}) ,  \nn \\
&=\  q\gamma^1 \gamma^0 \cosh\xi\Psi_{(p+1)(q-1)}  +q \gamma^1\gamma^2 \sinh\xi\Psi_{(p)(q)}.  
 \end{align}
 Again we have used the condition (\ref{tr2}) to obtain the last line. 
Substituting (\ref{az}), (\ref{ap}) and (\ref{am}) in (\ref{ga}) and using $\gamma^0\gamma^1\gamma^2=1=-\gamma^2\gamma^1\gamma^0$ we obtain.
\begin{align}\label{ga2}
 \Gamma^\mu \ma_{\mu(p)(q)} = - p\Psi_{(p-1)(q+1)} -q\Psi_{(p+1)(q-1)}. 
\end{align}
Thus equation (\ref{dirac-opr})  reduces to 
\begin{align}\label{ceq}
&\slashed{\nabla} \Psi_{(p)(q)} = \Delta \Psi_{(p)(q)} + p\Psi_{(p-1)(q+1)} + q\Psi_{(p+1)(q-1)},  \nn \\
\text{or \ \ }&\slashed{\nabla} \Psi_{(p)} = \Delta \Psi_{(p)}  + p\Psi_{(p-1)}+(s-p)\Psi_{(p+1)}.  
\end{align}
In the second line we have suppressed the label $(q)$ with the understanding that 
we will always have $p+q = n$. 
Thus we have reduced the action of the higher spin Dirac operator to that 
of the ordinary Dirac operator together with a mixing  `mass matrix'.

\subsection{Solutions of  the spin-$(n+\frac{1}{2})$ components}

Substituting (\ref{ceq}) into the Dirac equation equation (\ref{cs2}) we obtain
\begin{equation}
 \Delta \Psi_{(p)} +  p\Psi_{(p-1)} + (s-p)\Psi_{(p+1)} - m \Psi_{(p)} =0. 
\end{equation}
This can be written as
\begin{equation}
\label{coupl}
 \Delta \Psi_{(p)} - M^{(n)}_{pr} \Psi_{(r)} =0, 
\end{equation}
with the $(n+1)\times(n+1)$ matrix, $M_{pr}$ defined as
\begin{equation}
M^{(n)}_{pr} = -p\delta_{p-1,r} - (s-p)\delta_{p+1,r} + m\delta_{p,r}. 
\end{equation}
Note that this 
is a closed equation for the components of the tensor with boundary indices as $+$ or $-$.
It is basically $n+1$ coupled Dirac equations.  
In this section we will obtain the solutions for these components explicitly. 

\subsubsection*{Diagonalization of the mass matrix}

The first task  is 
decouple the equations in (\ref{coupl}) 
 or in other words diagonalize the matrix $M^{(n)}_{pr}$.
Following the method developed in \cite{Datta:2011za} we consider the linear
combination
\begin{equation}
\Psi_{[p]}=\sum_{a=0}^{p}\sum_{b=0}^{n-p} (-1)^b \binom{p}{a} \binom{n-p}{b} \Psi_{(n-a-b)}. 
\end{equation}
We can  also express $\Psi_{[p]}$ as, 
\begin{equation}\label{transformation}
\Psi_{[p]}=\sum _{q=0}^{s} T^{(n)}_{pq} \Psi_{(q)}.
\end{equation}
The transformation matrix  $T^{(n)}_{pq}$ is defined as follows. Consider the polynomial,
\begin{align}
 \sum_{q=0}^{n}T^{(n)}_{pq} x^q = \sum_{a=0}^{p}\sum_{b=0}^{n-p} (-1)^b \binom{n}{a} \binom{n-p}{b} x^{(n-a-b)}, 
\end{align}
this can be rewritten as
\begin{equation}
 \sum_{q=0}^{s}T^{(n)}_{pq} x^q = (x+1)^{p} (x-1)^{s-p}. 
\end{equation}
Thus
$T^{(s)}_{pq}$ is the coefficient of $x^q$ of the function above. 
A formal expression for $T^{(n)}_{pq}$ can then be obtained by a Taylor series expansion and be expressed as a contour integral.
\begin{equation}\label{T-expression}
T^{(n)}_{pq}=\frac{1}{2\pi i} \oint \frac{dx}{x^{q+1}} ( x+1)^{p} ( x-1)^{s-p}. 
\end{equation} 
The transformation matrix $T^{(n)}$ obeys  the following  identities

\vspace{.5cm}\noindent
\textbf{Identity 1} : 
\begin{equation}\label{claim-0}
\sum_{q=0}^{s} T^{(n)}_{pq}T^{(n)}_{qr} = 2^n \delta_{qr}. 
\end{equation}
This identity has been derived in Appendix C of \cite{Datta:2011za}. 
We also have 

\vspace{.5cm}\noindent
\textbf{Identity 2} :
\begin{equation}\label{claim-1}
\left( T^{(n)}M^{(n)}[T^{(n)}]^{-1} \right)_{pq}=(n-2p+m)\delta_{pq}.
\end{equation}
i.e., $T^{(n)}M^{(n)}[T^{(n)}]^{-1}$ is diagonal. 
The proof of this identity is provided in Appendix \ref{app-A}.
\def\p{{[p]}}
Substituting for $\Psi_{(p)}$ in terms of $\Psi_{[p]}$  and using 
(\ref{claim-1}) we obtain the following set of decoupled Dirac equations 
\begin{equation}\label{dirac-p}
\Delta \Psi _{[p]} - (m+n-2p)\Psi_{[p]} =0. 
\end{equation}
Since this is just a set of Dirac equations, they can be  easily solved using 
the solutions of \cite{Das:1999pt,Birmingham:2001pj}. 
We first substitute the following ansatz  for the two components of the spinor 
\def\aa{{\alpha '}}
\def\u{{(1)}}
\def\d{{(2)}}
\begin{equation} \label{full}
\Psi ^{(1,2)} _{[p]} = \frac{e^{-i(\kp x^+ + \km x^-)}}{\sqrt{\cosh\xi\sinh\xi}}  \psi ^{(1,2)} _{[p]} (\xi). 
\end{equation}
in the equation (\ref{dirac-p}).  
Note that with the definition of $x^+$ and $x^-$  given in (\ref{xpxm}) we see that the 
frequency and momenta of these solutions are related to $k_+$ and $k_-$ by 
\begin{equation}
 ( k_+ + k_-) ( r_+ - r_-) = \omega -k, \qquad ( k_+ -k_-) (r_+ + r_-) = \omega+k.
\end{equation}
Substituting (\ref{full})  into the decoupled Dirac equation (\ref{dirac-p})
 reduces the equation to 
\begin{equation}
\label{dirac-p1}
\gamma^1 \partial_\xi \psi_{[p]} - \gamma^0 \frac{ik_+}{\sinh\xi} \psi_{[p]} - \gamma^2 \frac{ik_-}{\cosh\xi}   \psi _{[p]}   - (m+n-2p)\psi_{[p]} =0. 
\end{equation}
We then  substitute 
\begin{equation} \label{plus-minus}
\psi^\pm_{[p]}   =\psi^\u_{[p]} \pm \psi^\d_{[p]} =(1-\tanh^2\xi)^{-1/4} \sqrt{1\pm\tanh\xi}\ (\psi'^\u_{[p]} \pm \psi'^\d_{[p]}). 
\end{equation}
in  (\ref{dirac-p1}) which leads to the following equations 
\begin{equation}
\begin{aligned}
2\sqrt{z}(1-z)\partial_z\psi'^\u_{[p]} + i\left(\frac{k_+}{\sqrt{z}} +k_- \sqrt{z}\right)\psi'^\u_{[p]} &= - \left[i(k_+ + k_-) -m-n+2p+\frac{1}{2}\right] \psi'^\d_{[p]},  \\
2\sqrt{z}(1-z)\partial_z\psi'^\d_{[p]} - i\left(\frac{k_+}{\sqrt{z}} +k_- \sqrt{z}\right)\psi'^\d_{[p]} &= - \left[-i(k_+ + k_-) -m-n+2p+\frac{1}{2}\right] \psi'^\u_{[p]}.  
\end{aligned}
\end{equation}
The solutions of these equations which obey the 
ingoing boundary conditions at the horizon are \cite{Birmingham:2001pj}
\begin{equation}
\begin{aligned}\label{sol}
\psi'^\u_{[p]} &= d_{[p]}z^\alpha (1-z)^{\beta_p} F(a_{[p]},b_{[p]},c_{[p]};z) , \\
\psi'^\d_{[p]} &= d_{[p]}\frac{a_{[p]}-c_{[p]}}{c_{[p]}} z^{\alpha+ 1/2}(1-z)^{\beta_{[p]}} F(a_{[p]},b_{[p]}+1,c_{[p]}+1;z) . 
\end{aligned}
\end{equation}
where $\alpha = -\frac{ik_+}{2}$, $\beta_{[p]}= \frac{1}{2}(m+n-2p-\frac{1}{2})$, $c_{[p]} = \frac{1}{2}+2\alpha$, and 
\begin{equation}
a_{[p]} = \frac{k_+ - k_-}{2i} + \beta_{[p]} + \frac{1}{2} \; , \quad b_{[p]} = \frac{k_+ + k_-}{2i} + \beta_{[p]}.
\end{equation}
Defining $e^{(1,2)}_{[p]}$ as,
\begin{equation} \label{ratio-orgin}
\begin{aligned}
e^\u_{[p]}&=d_{[p]. } \\
e^\d_{[p]}&=d_{[p]} \frac{a_\p-c_\p}{c_\p}. 
\end{aligned}
\end{equation}
the solutions (\ref{sol}) become
\begin{equation}
\begin{aligned}\label{sol2}
\psi'^\u_{[p]} &= e^\u_{[p]}z^\alpha (1-z)^{\beta_p} F(a_\p,b_\p,c_\p;z),  \\
\psi'^\d_{[p]} &= e^\d_{[p]} z^{\alpha+ 1/2}(1-z)^{\beta_\p} F(a_\p,b_\p+1,c_\p+1;z) . 
\end{aligned}
\end{equation}
Note that the  two components of  $\psi_{[p]}'$ are determined  completely up to 
the constants $e_{[p]}^{(1,2)}$ which we call the polarization constants. 

\subsection*{Behaviour of the solutions near the boundary}
\def\pp{{[p]}}
\def\po{{(p)}}

We shall take a look at the behaviour of the solutions near the boundary, \(z\rightarrow1\). This will enable us to fix the conformal dimension $\hat \Delta$ of the field. 

Expanding the solutions (\ref{sol2}) near \( z \rightarrow 1 \)
 and using (\ref{full}, \ref{plus-minus}) we have the following
 \begin{align}
\Psi^+_\pp &\sim \mathscr{C}_1 (1-z)^{ \frac{1}{2}-\frac{m}{2}- \frac{n}{2} +p },  \\
\Psi^-_\pp &\sim \mathscr{D}_1 (1-z)^{ 1-\frac{m}{2}- \frac{n}{2} +p } +\mathscr{D}_2 (1-z)^{ \frac{1}{2}+ \frac{m}{2}+ \frac{n}{2} +p }. 
 \end{align}
Here \(  \mathscr{C}_1, \ \mathscr{D}_1 \) and \(  \mathscr{D}_2 \) are constants. 
For definiteness we will take $m>0$.
Thus  the most singular behaviour near the boundary is given by 
\begin{eqnarray}
\Psi_{i_1 \cdots i _n}  &\sim& ( 1-z)^{\frac{1}{2} + \frac{m}{2} - \frac{n}{2} },  \\ \nonumber
&\sim&  \hat z ^{\delta} , \qquad r\rightarrow \infty,  
\end{eqnarray}
where
\begin{equation}
\delta = ( 1-m-n).  
\end{equation}
Here we have  re-written $z$ in terms of the  co-ordinate $\hat z$ 
which is related to the radial co-ordinate $r$ by 
\begin{equation}
\hat z = \frac{1}{r}.
\end{equation}
The reason for this is that asymptotically near the boundary the  BTZ metric (\ref{dbtz}) 
 reduces to the following  $AdS_3$ metric
\begin{equation}
\label{dbtz1}
ds^2 = \frac{1}{\hat z^2} ( -dt^2 + d \phi^2 + d\hat z^2) . 
\end{equation}
 $i_1, \cdots i_n$ 
denote the boundary co-ordinates. Note that using the condition in 
(\ref{tr2}) it is easy to see that that any component involving the radial 
co-ordinate is suppressed compared to the boundary components as
$r$ or $ \xi \rightarrow \infty$. 
The conformal dimension $\hat\Delta$ of the dual operator can then be 
obtained from the coupling of the boundary value of the spin $s$ field
to the corresponding operator 
which is given by 
\begin{equation}
\int d^2 x {\cal O }^{i_i, \cdots i_n} \Psi_{i_i, \cdots i_n} . 
\end{equation}
From conformal invariance we obtain the following expression for the conformal 
dimension of the dual operator
\begin{equation}
\label{mdim}
\hat \Delta = 2 - \delta - n = 1 +m. 
\end{equation}

\subsection*{Finding the polarization constants}

Our next task is to find the coefficients $e^{(1,2)}_{[p]}$. 
Note that from the definition of these constants in (\ref{ratio-orgin}) 
 we see that their ratio is given by 
 \begin{equation}
 \frac{ e_{[p]}^{(2)}}{
 e_{[p]}^{(1)} } = \frac{ a_{[p]} - c_{[p]}}{ c_{[p]} } 
 \end{equation}
 Therefore it is sufficient to determine $e_{[p]}^{(1)}$. 
 The Chern-Simons like equations in (\ref{hs-cs2}) relate the various polarization constants. 
 We will show that using the Chern-Simons equations and the condition (\ref{tr2}) it
 is possible to determine all the polarization constants in terms of  a single constant. 
 Since these are constants,
  it is sufficient to examine the equations and the functions
 near the horizon. 
To begin let us  recall the following relation 
\begin{equation}
\Psi^\u_\pp \pm \Psi^\d_\pp = \sqrt{\frac{\cosh\xi \pm \sinh\xi}{\cosh\xi \sinh\xi}} 
(\psi^{\prime\u}_\pp \pm \psi^{\prime \d}_\pp)e^{-i (k_+x^+ k_-x^-)}.
\end{equation}
Near the horizon \(z\rightarrow 0\) we have
\begin{equation}
 \sqrt{\frac{\cosh\xi \pm \sinh\xi}{\cosh\xi \sinh\xi}} \simeq \frac{1}{z^{1/4}} \left( 1 \pm \frac{\sqrt{z}}{2} + O(z) \right). 
\end{equation}
Thus the near horizon behaviours of the solutions are
\begin{align}
\Psi^\u_\pp &\simeq e^\u_\pp z^\aa e^{-i (k_+x^+ k_-x^-)},  \\
\Psi^\d_\pp &\simeq \lb \frac{e^\u_\pp}{2} +e^\d_\pp \rb z^{\aa +1/2}e^{-i (k_+x^+ k_-x^-)}. 
\end{align}
here \([p]=0, 1,2,\cdots,n\). Note that 
\( \Psi^{(1,2)}_{\po} \) is a linear combination
of $\Psi^{(1,2)}_{[p]}$   which is given in (\ref{transformation}).
 This implies that the  behaviour of \( \Psi^{(1,2)}_{\po} \) near the horizon
 $z\rightarrow 0$  are given by
\begin{equation}
\label{lead1}
\Psi_\po^\u \simeq e_\po^\u z^\aa e^{-i (k_+x^+ k_-x^-)}, \qquad 
\Psi_\po^\d \simeq e_\po^\d z^{\aa +1/2}e^{-i (k_+x^+ k_-x^-)} . 
\end{equation}
with 
\begin{equation}
\label{poltran}
e_{[p]}^{(1)}=\sum _{q=0}^{s} T^{(n)}_{pq} e_{(q)}^{(1)}, 
\end{equation}
and a similar relation for the second component of the polarization. 
We will see subsequently that we will obtain closed equations for the 
polarization $e_{(p)}^{(1)}$ which will be sufficient to obtain all the 
polarization constants in terms of a single one.  
In order to obtain 
the behaviour of  $\Psi^{(1,2)}_{\xi(p)(q)}$ we  consider the tracelessness condition 
\begin{equation}
\Gamma ^\mu \Psi _{\mu(p)(q)} = 0. 
\end{equation}
This results in 
\begin{equation}
\Psi _{\xi(p)(q)} = - \frac{1}{\sinh\xi} \gamma^1 \gamma^0 \Psi _{+(p)(q)}  - \frac{1}{\cosh\xi} \gamma^1 \gamma^2 \Psi _{-(p)(q)} , 
\end{equation}
From examining the leading terms in the above 
equation near the horizon, $z\rightarrow 0$ 
we obtain
\begin{equation}
\Psi _{\xi(p)(q)} \simeq \begin{pmatrix}
e^\u_{+(p)(q)} z^{\aa -1/2} e^{-i (k_+x^+ k_-x^-)} \\
-( e^\d_{+(p)(q)}  +  e^\u_{-(p)(q)} ) z^\aa e^{-i (k_+x^+ k_-x^-)}
\end{pmatrix}. 
\end{equation}
Thus the behaviour of  $\Psi^{(1,2)}_{\xi(p)(q)}$ near the horizon  is
\begin{equation} \label{lead2}
\Psi^\u_{\xi(p)(q)}\simeq e^\u_{\xi(p)(q)} z^{\aa -1/2}e^{-i (k_+x^+ k_-x^-)}, \qquad 
\Psi^\d_{\xi(p)(q)}\simeq e^\d_{\xi(p)(q)} z^{\aa}e^{-i (k_+x^+ k_-x^-)} .
\end{equation}
where
\begin{equation}\label{coe-1}
e^\u_{\xi(p)(q)}=e^\u_{+(p)(q)}, \qquad e^\d_{\xi(p)(q)}=-( e^\d_{+(p)(q)}  +  e^\u_{-(p)(q)} ).  
\end{equation}

To obtain a closed set of equations for $e^{(1)}_{\pm(p)(q)}$. we will need to find the 
relation between  $e^{(1)}_{\xi(p)(q)}$ and  $e^{(1)}_{\pm(p)(q)}$.  
For this we consider the `$-(p)(q)$' component of the Chern-Simons  equation
\begin{equation}
\Gamma_-^{\ \nu\rho} \nabla _\nu \Psi_{\rho(p)(q)} = m\Psi_{-(p)(q)}. 
\end{equation}
Expanding this equation and rearranging terms we obtain
\begin{align}
\partial_+ \Psi_{\xi(p)(q)} - 2\sqrt{z}(1-z)\partial_z \Psi_{(p+1)(q)} + p\sqrt{z} \Psi_{(p-1)(q+2)} &+ q\sqrt{z} \Psi_{(p+1)(q)}  \nn \\
 - \frac{1}{2\sqrt{1-z}}\sigma^{01}\Psi_{\xi(p)(q)} &= m\sqrt{z}\Psi_{(p)(q+1)}. 
\end{align}
The  tracelessness condition  (\ref{double-trace}) was used to simplify the above equation. Near the horizon the leading terms in the above equation reduces to 
\begin{align}\label{tr-horizon}
\partial_+ \Psi_{\xi(p)(q)} - 2\sqrt{z}(1-z)\partial_z \Psi_{(p+1)(q)} + p\sqrt{z} \Psi_{(p-1)(q+2)} &+ q\sqrt{z} \Psi_{(p+1)(q)}  \nn \\
 - \frac{1}{2}\sigma^{01}\Psi_{\xi(p)(q)} &= m\sqrt{z}\Psi_{(p)(q+1)}. 
\end{align}
We now examine the 
`1' component of the above   spinor equation.  
Using (\ref{lead1}), (\ref{lead2}) we see that 
 near the horizon $z\rightarrow 0$, 
the  leading terms in the above equation is of the order $z^{\alpha-\frac{1}{2}}$.
Since the equation must hold to the leading order we have the equality
 \begin{equation}
-ik_+ e^{\u}_{\xi(p)(q)}z^{\aa-1/2} - 2\sqrt{z}\partial_z(e^{\u}_{(p+1)(q)}z^\aa )
=0. 
\end{equation}
This results in the following relation
\begin{equation}\label{coe-relation}
e^\u_{\xi(p)(q)} = e^\u_{+(p)(q)}. 
\end{equation}
We now have sufficient information to find the closed set of equations  for the $ e^{(1)}_{(p)}$. We consider the `$\xi(p)(q)$' component of the Chern-Simons equation
\begin{equation}
g_{\xi\xi}\Gamma^{\xi+-} (\nabla_+ \Psi_{-(p)(q)} - \nabla_- \Psi_{+(p)(q)})= m\Psi_{\xi(p)(q)} . 
\end{equation}
Writing this explicitly by expanding each term we have
\begin{align}\label{pre-rec}
&\partial_+\Psi_{-(p)(q)} -\frac{1}{2}\cosh\xi \sigma^{01}\Psi_{-(p)(q)} - \frac{p\sqrt{z}}{1-z} \Psi_{\xi-(p-1)(q)} \nn \\
&-\partial_-\Psi_{+(p)(q)} - \frac{1}{2}\sinh\xi \sigma^{21}\Psi_{+(p)(q)} -\frac{q\sqrt{z}}{1-z}\Psi_{\xi+(p)(q-1)}  = -\frac{m\sqrt{z}}{1-z}\Psi_{\xi(p)(q)}.  
\end{align}
Examining the `1' component of the equation in (\ref{pre-rec}) near horizon $z\rightarrow 0$,
we see that the leading terms go as $\sim z^{\alpha}$. Requiring the leading terms
to satisfy the equation in (\ref{pre-rec}) we obtain 
\begin{eqnarray}
&& -i\kp e^\u _{-(p)(q)} z^\aa -\frac{1}{2}e^\u _{-(p)(q)}z^\aa - p e^\u_{\xi(p-1)(q+1)}z^\aa  \\ \nonumber
&& \qquad \qquad\qquad +i\km e^\u _{+(p)(q)}z^\aa
- q e^\u_{\xi(p+1)(q-1)}z^\aa  = -me^\u_{\xi(p)(q)}  z^{\aa } .
\end{eqnarray}
Now using the (\ref{coe-relation})  and $q = n -p-1$ we obtain the recursion relations
\begin{equation}\label{rec-1}
(n-p-1)e^\u_{(p+2)} + (-m-i\km)e^\u_{(p+1)} + (i\kp +\frac{1}{2} + p)e^\u_{(p)}=0. 
\end{equation}
These are a set of $n$ equations for the $n+1$ variables. 
Let us define the `recursion matrix' $\tilde C_{ij}^{(n)}$ as
\begin{equation}
\tilde C_{jl}^{(n)} = ( n -j-1) \delta_{j+2, l} + ( -m -ik_-)\delta_{j+1, l} + ( j + i k_+ + \frac{1}{2} ) \delta_{j, l}.  
\end{equation}
Note that $j$ runs from $n-1$ to $0$ and $l$ runs from $n$ to $0$. 
Thus we can write the recursion relations in (\ref{rec-1}) as 
\begin{equation}
\label{rec1}
\sum_{l =0}^{n-1} \tilde C_{jl}^{(s)} e_{(l)}^{(1)} = 0 , \qquad \mbox{for} \qquad j =  0, 1, 2, \ldots , n-1. 
\end{equation}
This is same recursion  relation for the polarization coefficients obtained in \cite{Datta:2011za} for the higher spin bosonic case with following 
 replacements 
\begin{equation}
\label{repl}
 s\rightarrow n, \qquad   m\rightarrow -m, \qquad i\kp \rightarrow i\kp +\frac{1}{2}.  
\end{equation} 
 From \cite{Datta:2011za} we see that the recursion relation is easily  solved by the change 
 of basis to the polarization coefficients $e^\u_{[p]}$. 
 That is we have the identity 
 
 \vspace{.5cm}\noindent
 {\bf Identity 3:}
 \vspace{.5cm}
 \begin{eqnarray}
&& \frac{1}{2^{n-1}} ( T^{(n-1)} \tilde 
 C^{(n)} T^{(n)} )_{jl} \\ \nonumber 
&&= ( 2j - n +\frac{3}{2}  -m + i (k_+ -k_-) ) \delta_{j+1, l }
 + ( 2j - n +\frac{1}{2} -m - i( k_+ + k_-) ) \delta_{j, l }. 
 \end{eqnarray}
 \vspace{.5cm}
 
 \noindent
The proof of 
this identity can be obtained using the same method as that of Identity 3 in \cite{Datta:2011za}
but with the replacements given in (\ref{repl}). 
 Now performing the change of basis  in the recursion relations given in 
 (\ref{rec1}) using  the transformation in (\ref{poltran}) we obtain 
\begin{align}
(2j-n+\frac{3}{2}-m+i(\kp-\km))e^\u_{[j+1]} + (2j-n+\frac{1}{2}-m-i(\kp+\km))e^\u_{[j]} =0 . 
\end{align}
From this we get
\begin{equation}
e^\u_{[n-j]} = - \frac{2j-n-\frac{3}{2}+m-i(\kp-\km)}{2j-n-\frac{1}{2}+m+i(\kp+\km)} e^\u_{[n-j+1]}. 
\end{equation}
One can then write all the coefficients $e^\u_{[n-j]}$ in terms of $e^\u_{[n]}$.
\begin{equation}\label{e1}
e^\u_{[n-j]} = (-1)^j \prod^{j-1}_{u=0}  \frac{2j-n+\frac{1}{2}+m-i(\kp-\km)}{2j-n+\frac{3}{2}+m+i(\kp+\km)} e^\u_{[n]}. 
\end{equation}
We can now solve for  the polarization components $e^{(2)}_{[n-j]}$.
From  (\ref{ratio-orgin}) which determines the ratio 
between the \(e^\d_{[n-j]}\)  and \(e^\u_{[n-j]}\) we obtain
\begin{equation}
\label{rr1}
\frac{e^\d_{[n-j]}}{e^\u_{[n-j]}}  = \frac{a_{[n-j]}-c_{[n-j]}}{c_{[n-j]}}  =   \frac{2j-n+m-\frac{1}{2}+i(\kp + \km)}{\frac{1}{2}-i\kp}.
\end{equation}
Substituting  (\ref{e1}) in the above equation we obtain
\begin{equation}\label{e2}
e^\d_{[n-j]} = (-1)^j \prod^{j-1}_{u=0}  \frac{2u-n+\frac{1}{2}+m-i(\kp-\km)}{2u-n-\frac{1}{2}+m+i(\kp+\km)} e^\d_{[n]}.
\end{equation}
Note that $e^\d_{[n]}$ is also determined in terms of $e^\u_{[n]}$ by 
(\ref{rr1}). Thus all polarization constants are determined in terms of a single constant.

\subsection{Quasinormal modes}

We can now substitute the values of the polarization constants and obtain 
the final form of the 
the solutions \(\psi'^\u_{[u-j]} \)  and \( \psi'^\d_{[u-j]} \). 
These are given by 
\begin{align}\label{sol3}
\psi'^\u_{[n-j]} &=  (-1)^j \prod^{j-1}_{u=0}  \frac{2u-n+\frac{1}{2}+m-i(\kp-\km)}{2u-n+\frac{3}{2}+m+i(\kp+\km)} e^\u_{[n]} \nn \\
& \qquad\quad \times z^\alpha (1-z)^{\beta_{[n-j]} } F(a_{[n-j]},b_{[n-j]},c_{[n-j]};z),  \\
\psi'^\d_{[n-j]} &=  (-1)^j \prod^{j-1}_{u=0}  \frac{2u-n+\frac{1}{2}+m-i(\kp-\km)}{2u-n-\frac{1}{2}+m+i(\kp+\km)} e^\d_{[n]} \nn \\
& \qquad\quad \times z^{\alpha+ 1/2}(1-z)^{\beta_{[n-j]} } F(a_{[n-j]},b_{[n-j]}+1,c_{[n-j]}+1;z) . 
\end{align}
To obtain the quasi-normal modes we need to impose the vanishing 
Dirichlet condition at the boundary $z\rightarrow 1$. For the case $m>0$, the dominant
behaviour of the solutions near the boundary is given by 
\begin{align}\label{sol4}
\psi'^\u_{[n-j]} &\simeq  (-1)^j \prod^{j-1}_{u=0}  \frac{2u-n+\frac{1}{2}+m-i(\kp-\km)}{2u-n+\frac{3}{2}+m+i(\kp+\km)} e^\u_{[n]} \nn \\
& \qquad\quad \times  (1-z)^{-\beta_{[n-j]}  } \frac{\Gamma(c_{[n-j]})\Gamma(a_{[n-j]}+b_{[n-j]}-c_{[n-j]})}{\Gamma(a_{n-j})\Gamma(b_{n-j})} , \\
\psi'^\d_{n-j} &\simeq  (-1)^j \prod^{j-1}_{u=0}  \frac{2u-n+\frac{1}{2}+m-i(\kp-\km)}{2u-n-\frac{1}{2}+m+i(\kp+\km)} e^\d_{[n]} \nn \\
& \qquad\quad \times(1-z)^{-\beta_{n-j} } \frac{\Gamma(c_{[n-j]}+1)\Gamma(a_{[n-j]}+b_{[n-j]}-c_{[n-j]})}{\Gamma(a_{[n-j]})\Gamma(b_{[n-j]}+1)} .  \label{sol42}
\end{align}
We can obtain the 
 quasinormal modes by requiring that the coefficients of these 
leading terms vanish.   Note that
\begin{equation}
a_{[n-j]} = a_{[n]} + j, 
\end{equation}
From this we have
\begin{align}
\Gamma(a_{[n-j]}) &= \Gamma(a_{[n]} + j),  \nn \\
&=  \Gamma(a_{[n]}) \prod^{j-1}_{u=0} (a_{[n]} + u),  \nn \\
&= \frac{ \Gamma(a_{[n]})}{2^j} \prod^{j-1}_{u=0} ( 2u-n+\frac{1}{2}+m-i(\kp-\km)  ). 
\end{align}
The product $ \prod_{u=0}^{j-1} ( 2u-n+\frac{1}{2}+m-i(\kp-\km)  )$ 
exactly cancels the numerator of the coefficient in (\ref{sol4}) and (\ref{sol42}). 
Thus the behaviour of the solutions near the boundary reduces to 
\begin{align}\label{sol51}
\psi'^\u_{[n-j]} &\simeq  \frac{(-2)^j}{  \prod^{j-1}_{u=0} 2u-n+\frac{3}{2}+m+i(\kp+\km)} e^\u_{[n]} \nn \\
& \qquad\quad \times  (1-z)^{-\beta_{[n-j]}} \frac{\Gamma(c_{[n-j]})\Gamma(a_{[n-j]}+b_{[n-j]}-c_{[n-j]})}{\Gamma(a_{[n]})\Gamma(b_{[n-j]})},  \\
\psi'^\d_{[n-j]} &\simeq  \frac{(-2)^j}{ \prod^{j-1}_{u=0} 2u-n-\frac{1}{2}+m+i(\kp+\km)} e^\d_{[n]} \nn \\
& \qquad\quad \times(1-z)^{-\beta_{[n-j]}} \frac{\Gamma(c_{[n-j]}+1)\Gamma(a_{[n-j]}+b_{[n-j]}-c_{[n-j]})}{\Gamma(a_{[n]})\Gamma(b_{[n-j]}+1)}.   \label{sol52}
\end{align}
The vanishing Dirichlet conditions at the boundary constrain 
the leading behaviour of all components of the tensorial spinor to  vanish at infinity. 
As argued below (\ref{dbtz1}) the components which involve the radial coordinate are suppressed 
at the boundary compared to the components which only involved the boundary coordinates $+, -$. 
Thus it is sufficient to impose vanishing  Dirichlet  boundary conditions on these components. 
This is equivalent to looking at the common set of zeroes of the coefficients which 
occur in the functions given in (\ref{sol51}). These are clearly given by
\begin{align}
a_{[n]}=-\hat n \quad \text{and} \quad b_{[0]} +1=-\hat n , \qquad \hat n =0,1,2,\ldots. 
\end{align}
In terms of $\kp$ and $\km$ these translate to
\begin{equation}
i(\kp-\km) =2\hat n  +\hat\Delta-s , \qquad i(\kp+\km)=2\hat n  + \hat\Delta+s, \qquad \hat n =0,1,2,\ldots, 
\end{equation}
where $s=n+\frac{1}{2}$. Thus the quasinormal modes are given by
\begin{equation}
\begin{aligned}
\omega _L &=  k+2\pi T_L(k_+ +k_-),  \\
&=  k-2\pi iT_L(2 \hat n + \hat \Delta +s),
\end{aligned}\qquad
\begin{aligned}
\omega _R &= -k+2\pi T_R(k_+ -k_-),  \\ 
&= -k-2\pi iT_R (2 \hat n +\hat \Delta -s).
\end{aligned}
\end{equation}
These coincide precisely with the poles of the corresponding two  point function (\ref{locp}) for the 
corresponding spin $s$ field as expected from the AdS/CFT correspondence. 
Reading out $h_L$ and $h_R$ we get $h_R -h_L=-s$, the case  $h_R-h_L =+s $ will arise when 
we carry out the same analysis but with $m<0$. 
\def\ns{\slashed{\nabla}}
\def\mt{{\mathscr{M}}_s} 
\def\pol{\text{Pol($\hat \Delta$)}}
\def\bz{\bar{z}}

\section{1-loop determinant for arbitrary half-integer spins}

As shown in \cite{Denef:2009kn}, the poles of the retarded Green's function can be  used to 
construct the one-loop determinant of the corresponding field in the bulk. 
In \cite{Denef:2009kn}, the one-loop determinant for scalars in asymptotically $AdS$ black holes including the 
BTZ black holes was constructed using analyticity and the information of the quasinormal modes. 
This construction was extended to arbitrary integer spins  in \cite{Datta:2011za}. 
In this section we would like to use the quasinormal mode of the higher spin 
fermion along with analyticity to construct the one loop determinant of the 
corresponding half integer spin field. 
We will then show that this determinant agrees with that constructed in \cite{David:2009xg}
using group theoretic methods.

\subsection{1-loop determinant from the spectrum of quasinormal modes}

Our analysis will follow the method developed in \cite{Denef:2009kn,David:2009xg}. 
We consider the non-rotating BTZ black hole for which the metric is given by
\begin{equation}
 ds^2 = -( r^2 - r_+^2) dt^2 + \frac{dr^2}{r^2 - r_+^2} + r^2 d\phi^2. 
\end{equation}
We then continue the BTZ black hole to Euclidean time together with the identification
\begin{equation}
\label{period}
 t = - i\tau, \qquad \tau \sim \tau + \frac{1}{T}, 
\end{equation}
and 
\begin{equation}
 T = T_H = T_L = \frac{r_+}{2\pi}. 
\end{equation}
Our goal to evaluate the following one loop determinant of the higher spin  $s = n + \frac{1}{2}$
Laplacian 
\begin{equation} \label{defdet}
 Z_s( \hat \Delta) = {\rm det}  ( - \nabla^2(s) +\mt^2),  
\end{equation}
where $\mt$ is  the mass shift which occurs on squaring the Dirac operator 
which acts on the higher spin fermion. 
The result of  squaring the Dirac operator is given by 
\begin{equation}\label{2nd-o-1}
(\ns + m_n)(\ns -m_n)\Psi_{\mu_1\mu_2\dots\mu_n}=\left(\nabla^2 + (s+1) - m_n^2  \right) \Psi_{\mu_1\mu_2\dots\mu_n}=0.
\end{equation}
where $s=n+\frac{1}{2}$. 
The proof of the above identity is given in Appendix A. 
Thus the mass  $\mt$ is given by 
\begin{equation}\label{m-ss}
\mt^2 = m_n^2 - (s+1). 
\end{equation}
The basic strategy adopted by \cite{Denef:2009kn} to evaluate the determinant  in 
(\ref{defdet}) 
 is to identify the zeros of the determinant in the complex $\hat \Delta$ space. 
Here $\hat \Delta$ is the conformal dimension of the corresponding dual operator. 
This occurs whenever the wave equation of the corresponding field has a zero
mode and also obeys the periodicity (\ref{period}).  For the case of the spin $s$ field 
these  modes are given by 
\begin{equation} 
\label{qnmo}
\omega_L \equiv \omega_L  = k -2\pi i T ( 2\hat n + \hat\Delta +s) , \qquad
\omega_R  \equiv \omega_R = -k - 2\pi i T( 2\hat n  +\hat \Delta -s). 
\end{equation}
where \( \hat n  =0,1,2,\dots\) . 
where $k$ is the momentum along the $\phi$ direction. 
Thus it is quantized and therefore
takes values in the set of integers
\begin{equation}
 k = 0, \pm 1, \pm2, \ldots. 
\end{equation}
Note that for the modes in (\ref{qnmo}) we have considered the situation when $h_L > h_R$. 
We now define
\begin{align}
z_L &= k - 2\pi i T( 2n + \hat \Delta +s), \quad   && \bar z_L = k + 2\pi i T( 2n + \hat \Delta +s) , 
\\ \nonumber
z_R &= -k - 2\pi i T( 2n + \hat \Delta - s), \quad   && \bar z_R = -k + 2\pi i T( 2n + \hat \Delta -s) .
\end{align}
Requiring the quasinormal modes to obey the thermal periodicity conditions due to the 
identification in (\ref{period}) results in the following equations 
\begin{align}\label{qcond}
2\pi i T(\tilde n    - s) &= z_L(\hat \Delta) , \qquad \tilde n \geq  0 , \\ \nonumber
2\pi i T( \tilde n   + s) &= \bar z_L(\hat\Delta), \qquad \tilde n <0, \\ \nonumber
2\pi i T( \tilde n    +s) &= z_R( \hat\Delta), \qquad \tilde n  \geq 0, \\ \nonumber
2\pi i T( \tilde n   - s) &= \bar z_R( \hat\Delta), \qquad \tilde n <0. 
\end{align}
Note that since  $\tilde n$ is an integer $\frac{  z_{L,R} (\hat \Delta)}{2\pi i T}$ is half integral moded. 
These equalities imply that when $\hat\Delta$ is tuned to these values, the one
loop determinant exhibits zeros. The ranges $\tilde n $ are chosen so that the quantities
\begin{equation}
k - 2\pi i T( 2  n + \hat \Delta) , \qquad \hbox{and} \quad k + 2\pi i T ( 2n + \hat \Delta). 
\end{equation}
when considered together take values $2\pi i T  \tilde n  $ where $\tilde n$ assumes 
values in the set of integers. Similarly the range of $\tilde n$ for the case of the right-moving
quasinormal modes is chosen so that the quantities 
\begin{equation}
-k - 2\pi i T( 2  n + \hat \Delta) , \qquad \hbox{and} \quad - k + 2\pi i T ( 2n + \hat \Delta). 
\end{equation}
when considered together take values $2\pi  i T  \tilde n   $ 
where $\tilde n$ assumes values in the set of integers.  Note that these 
are the same quantization conditions used in \cite{Datta:2011za}  for the case of 
integer spin $s$. 
Though we do not have a first principle justification of the choice of these ranges we will 
show that they do indeed lead to the answer evaluated using the group theoretic methods 
given in \cite{David:2009xg}. 
The function which is analytic in $\hat\Delta$ and has zeros at the locations (\ref{qcond}) is 
given  by
\begin{align}\label{pf}
Z_{(s)} =  \   e^{\pol} \prod _{\substack{z_L,\bz_L \\ z_R,\bz_R}} \Bigg{[} &  
\left( -s+\frac{iz_L}{2\pi T} \right)^{} 
\left( s+\frac{iz_R}{2\pi T} \right)^{}
 \nonumber \\
\times   &\; \prod _{v+ \frac{1}{2} > -s} \left( v+\frac{1}{2} + \frac{iz_L}{2\pi T} \right)^{}  \left( v+ \frac{1}{2} -\frac{i\bz_L}{2\pi T} \right)^{} \nn \\
\times   &\;\;  \prod _{v+ \frac{1}{2} > s} \left( v+\frac{1}{2} + \frac{iz_R}{2\pi T} \right)^{}  \left( v + \frac{1}{2} -\frac{i\bz_R}{2\pi T} \right)^{}  \Bigg{]}.
\end{align}
where $\pol$ is a non-singular holomorphic function of $\hat\Delta$ and can be determined by 
examining the $\hat\Delta \rightarrow\infty $ behaviour. The product over 
$z_L, \bar z_L, z_R, \bar z_R$ mean the product over $k, n$ occurring in the definition of these 
variables. The variable $v$ takes values in the set of integers. 
Plugging in the quasi-normal modes into (\ref{pf}) and performing simple
manipulations we obtain 
\begin{align}
\label{loop1}
Z_{(s)} &=  \   e^{\pol} \, \prod_{N\geq 0, p} \Bigg{[}  \left(( 2N+\hat\Delta)^2 +\frac{p^2}{(2\pi T)^2} \right)^{-1}   \nn \\
&\qquad \qquad\qquad \qquad  \times \; \prod_{v\geq 0} \left((v+ 2N+\hat\Delta)^2 +\frac{p^2}{(2\pi T)^2} \right)^{2} \Bigg{]} .
\end{align}
The analysis here is for the case $h_L  - h_R=s$. On performing the evaluation for the one
loop determinant  for the situation 
with $h_R  -h_L =-s$ it can be shown that one obtains the same expression as in (\ref{loop1}). 
Since a higher spin fermion obeying the second order spin $s$ Laplacian contains 
both the modes  we need to take the square of the expression in (\ref{loop1}). 
Taking this into account and taking logarithms on both sides of the (\ref{loop1}) we obtain  
\begin{align}
 - \log Z_{(s)}  &=  - \pol - 4\sum_{v\geq 0, N\geq 0, p} \log \left((v+ 2N+\hat\Delta)^2 +\frac{p^2}{(2\pi T)^2} \right) \nn \\
&\qquad\qquad\quad + 2\sum _{N\geq 0,p} \log \left(( 2N+\hat\Delta)^2 +\frac{p^2}{(2\pi T)^2} \right)  \notag ,\\ 
	 &=  - \pol - 4 \sum_{v> 0, N\geq 0, p} \log \left((v+ 2N+\hat\Delta)^2 +\frac{p^2}{(2\pi T)^2} \right)  \nn \\
&\qquad\qquad\quad -  2\sum _{N\geq 0,p} \log \left(( 2N+\hat\Delta)^2 +\frac{p^2}{(2\pi T)^2} \right)  \notag ,\\
     &=  - \pol - 2 \sum _{\kappa \geq 0, p} (\kappa + 1) \log \left((\kappa +\hat\Delta)^2 +\frac{p^2}{(2\pi T)^2} \right).  
\end{align}
The sums over $v$ and $N$ were combined and written as a sum over $\kappa =n+2N$. We have also made use of $\log (a+ib) + \log (a-ib) = \log (a^2+b^2)$.
The divergent sums can then be extracted and absorbed in $\pol$. We then make use of the identity \(  \sum _{p\geq1} \log \left( 1+ \frac{x^2}{p^2}\right) = \log \frac{\sinh  \pi x}{\pi x} = \pi x -\log (\pi x) + \log (1-2 e^{-2\pi x}) \) to  obtain 
\begin{equation}\label{pf3}
\log Z_{(s)} = \pol - 4 \log \prod _{\kappa \geq 0} (1- q^{-\kappa +\hat\Delta})^{-(\kappa + 1 )},
\end{equation}
where,
\begin{equation}
q=e^{2\pi i \tilde{\tau}} \ , \quad \quad \tilde{\tau}=2\pi i T . 
\end{equation}
To determine $\pol$ we use the same argument as in \cite{Denef:2009kn}. Note that 
taking the $\hat \Delta \rightarrow \infty$ the partition function should reduce to that of the 
BTZ which is locally identical to that of $AdS_3$. This determines $\pol$ to be a function 
proportional to the volume of the Euclidean BTZ black hole. We will not write this explicitly
since we do not require it in the subsequent discussion. 

\def\half{\frac{1}{2}}
\subsection{1-loop determinant from the heat kernel}

We now show that the finite term in the one loop partition function (\ref{pf3})
which is determined 
from the product over quasinormal modes agrees with that constructed from the 
heat kernel of the spin $s$ field.  
The trace of the heat kernel for the spin $s$ Laplacian on thermal 
$AdS_3$ is given by \cite{ David:2009xg}
\begin{equation}
{\rm Tr} ( e^{-t \nabla^2_{(s)}}) = 
K^{(s)}(\tau,\bar{\tau};t) = \sum ^{\infty} _{n=1} \frac{\tau _2}{\sqrt{4\pi t}|\sin \frac{n\tau}{2}|^2}\cos (sn\tau_1) e^{-\frac{n^2 \tau _2^2}{4t}} e^{-(s+1)t}.
\end{equation}
The formula retains only the finite term in the heat kernel and suppresses the term
which is proportional to the volume of the $AdS_3$. The parameter $\tau$ is related to the 
temperature of the Euclidean non-rotating BTZ by 
\begin{equation}
\tau = \frac{i}{2\pi T}.
\end{equation}
We will substitute this value of $\tau$ at the end of our analysis. 
The 1-loop determinant is then given by 
\begin{equation}
- \log (\det (-\nabla ^2 + \mt^2 )) = \int _0^{\infty} \frac{dt}{t} e^{-\mt ^2 t} K^{(s)}(\tau,\bar{\tau};t) ,
\end{equation}
where $\mt$ is given in (\ref{m-ss}). Substituting the value of $\mt$  we obtain 
\begin{equation}
e^{-\mt ^2 t} K^{(s)} (\tau,\bar{\tau};t)=\sum _{u=1} ^\infty \frac{\tau _2}{\sqrt{4\pi t}|\sin \frac{u\tau}{2}|^2}\cos (su\tau_1) e^{-\frac{u^2\tau_2^{\ 2}}{4t}}          e^{-m_n^2t}.
\end{equation}    
The integration over $t$ can be performed as follows
\begin{equation}
\frac{1}{\sqrt{4\pi}} \int_0^\infty \frac{\,dt}{t^{3/2}} e^{-\frac{u^2\tau^2_2}{4t}}e^{-m_n^2t}=\frac{1}{u\tau_2}
 e^{-u\tau_2 m_n } = e^{-u\tau_2(\hat\Delta -1 )}.
\end{equation}
Here we have used the relation $\hat \Delta = 1+ m=1+m_n$ derived in (\ref{mdim}). 
\def\bq{\bar{q}}
Thus the one loop determinant reduces to 
\begin{equation}
\begin{aligned}\label{q3m}
- \log (\det (-\nabla^2+\mt^2)) &= \sum _{u=1} ^\infty \frac{\cos (su\tau_1)}{u|\sin \frac{u\tau}{2}|^2}  e^{-u\tau_2(\hat\Delta -1)}, \\
&= \sum _{u=1} ^\infty \frac{2}{u} \frac{(q^{su}+\bq^{su})}{|1-q^u|^2}q^{(\hat\Delta-s)u} ,\\
&= \sum _{u=1} ^\infty \frac{4}{u} \frac{q^{\hat\Delta u}}{(1-q^u)^2}, \\
&= -4 \log \prod_{v=0} ^\infty (1-q^{v+\hat\Delta} )^{m+1} .
\end{aligned} 
\end{equation}
where 
\begin{equation}
q = e^{2\pi i \tau}.
\end{equation}
In the above manipulations we have also used the fact that $\tau$ is purely imaginary for the case of the 
non-rotating BTZ black hole which results in $q=\bar q$. 
Thus we obtain
\begin{equation}\label{qf1/2}
 \log Z_{(s)}= \log (\det (-\nabla^2+\mt^2)) = 4 \log \prod_{v=0} ^\infty (1-q^{v+\hat\Delta} )^{-(m+1)}.
\end{equation}
Comparing (\ref{pf3}) and (\ref{qf1/2}) we see that the two expressions indeed 
agree on performing the modular transformation
\begin{equation}
\tilde \tau  = - \frac{1}{\tau}.
\end{equation}
which is the expected relation between one-loop determinants on Euclidean BTZ and thermal 
$AdS_3$. 

\section{Conclusions}

We have solved the wave equations for arbitrary massive higher spin fermionic fields
in the BTZ background. In this work we focused on the ingoing modes at the 
horizon to obtain the
quasi-normal modes, but the analysis can be easily extended for the outgoing modes. 
This will lead to the complete set of modes for the higher spin fermion in this background
which is the starting point for its quantization. 
This can be useful for studying fermionic emission by Hawking radiation on similar lines
as \cite{Dasgupta:1998jg}. 
It will also be useful to find the exact prescription to evaluate the retarded Green's function 
for higher spin fermions  extending the work done for the spin $1/2$ and spin $3/2$ cases
in \cite{Iqbal:2009fd} and \cite{Gauntlett:2011wm} respectively. 

From our discussion of the wave equations for 
massive higher spin fermionic fields  it is clear that other properties like the 
bulk to boundary propagator for these fields can also be solved  and obtained in closed form. 
These are important tools to study the general $AdS_3/CFT_2$ correspondence 
and it is useful to obtain them. 
Finally the work in this paper and that related to massive higher spin integer spins 
in \cite{Datta:2011za} and the observation that classical string propagation in 
 BTZ is integrable \cite{David:2011iy} suggests that it is possible to quantize 
strings in the BTZ background. 

\acknowledgments

J.R.D thanks the organizers of the `The Third Indian-Israeli International 
meeting on  String Theory: Holography and its applications'
for the stimulating meeting  and for an opportunity to present this work. 
The work of J.R.D is partially supported by the 
Ramanujan fellowship DST-SR/S2/RJN-59/2009. 

\appendix
\def\sg{\sqrt{-g}}
\def\pd{\partial}
\section{The background geometry }

\subsection*{The BTZ black hole}

The metric of the BTZ black hole is conventionally written as
\begin{eqnarray}
\label{metconv}
 ds^2 &=&  -\frac{\Delta^2}{ r^2} dt^2 + \frac{r^2}{\Delta^2} dr^2 + 
r^2 \left( d\phi - \frac{r_+r_-}{r^2} dt\right)^2, \\ \nonumber
\Delta^2 & = & ( r^2 - r_+^2) ( r^2 - r_-^2).  
\end{eqnarray}
Here $r_+$ and $r_-$ are the radii of the inner and outer horizons respectively,  
$r$ is the radial distance and $t$ labels the time. The angular coordinate
$\phi$ has the period of $2\pi$.   
We will work with units in which  the radius
of $AdS_3$ is unity. 
 The left and right temperatures are defined as 
\begin{equation}
 T_L = \frac{1}{2\pi} ( r_+ - r_-) , \qquad T_R = \frac{1}{2\pi} ( r_+ + r_-) .
\end{equation}
A convenient coordinate system for our analysis was discovered by \cite{Birmingham:2001pj}. We first define the coordinates
\begin{align}
\label{xpxm}
z = \tanh^2 \xi  &= \frac{r^2 - r_+^2}{r^2 - r_-^2}, \\ \nonumber
 x^+ = r_+ t - r_- \phi,& \qquad x^- = r_+ \phi - r_- t.
\end{align}
Note that in these coordinates, the range of $r$ from $r_+$ to $\infty$ is mapped to 
$z=0$ or $\xi =0$ to $z=1$ or $\xi = \infty$ respectively. 
In these coordinates, the BTZ metric given in (\ref{metconv})  reduces to  the following diagonal metric
\begin{align}
\label{diamet}
ds^2 &= d\xi ^2 - \sinh^2 \xi \, dx_+^2 + \cosh^2 \xi \, dx_-^2 .
\end{align}
This form of the metric is used in our calculations and we will briefly list its various
properties.
The non-vanishing Christoffel  symbols of the metric in (\ref{diamet}) are  given by
\begin{eqnarray}
\begin{aligned}
\label{chris}
 &\tilde \Gamma_{++}^\xi = \cosh\xi \sinh\xi =\frac{\sqrt{z}}{1-z}, \quad &&\tilde \Gamma^\xi_{--} = - \cosh\xi \sinh \xi = -\frac{\sqrt{z}}{1-z} , \\ \nonumber
 &\tilde \Gamma^{+}_{+\xi} = \coth \xi=\frac{1}{\sqrt{z}}, \quad &&
 \tilde \Gamma^{-}_{- \xi} = \tanh\xi=\sqrt{z} \ . 
\end{aligned}
\end{eqnarray}
The metric and its Christoffel symbols obey the following identities
which will be useful in simplifying the higher spin equations in the next sections 
\begin{eqnarray}
\sg &=&  \cosh \xi \sinh \xi = \frac{\sqrt{z}}{1-z}\ ,\\
\frac{g_{++}}{\sg} &=& - \tanh \xi =-\sqrt{z} \ , \\
\frac{g_{--}}{\sg} &=&  \coth \xi =\frac{1}{\sqrt{z}} \ , \\
 \label{gdg} \frac{1}{\sg} \pd _\mu (\sg g^{\mu \nu} \tilde \Gamma ^{\sigma}_{\nu \rho}) 
&= & \label{c0}
\frac{1}{\sg} \pd _\xi (\sg g^{\xi\xi} \tilde\Gamma ^{\sigma}_{\xi \rho}) 
= 2 \hat\delta _{\rho}^{\sigma}  \\
\label{c1} g^{\pm \pm}\tilde \Gamma^\xi _{\pm \pm} + \tilde \Gamma ^\pm _{\xi\pm} &=&  0  \ , \\
g^{++}\tilde \Gamma^\xi _{++} &=&  - \coth \xi =-\frac{1}{\sqrt{z}}  \ , \\
g^{--}\tilde \Gamma^\xi _{--} &=&  - \tanh \xi =-\sqrt{z}  \ .
\end{eqnarray}
where $\hat \delta_{\rho\sigma}$ is defined as
\begin{equation}
\hat\delta_{\rho}^{\sigma} = 
\begin{cases}
1 & \text{for $\rho,\sigma = \pm$ and $\rho=\sigma$,}\\
0 & \text{otherwise.}
\end{cases}
\end{equation}

The BTZ black hole is obtained by 
identifications of $AdS_3$ \cite{Banados:1992gq}. 
Thus it is locally $AdS_3$ and therefore its curvature  obey the following relations
\begin{eqnarray}
\label{local}
R_{\alpha\beta\gamma\delta} &=& g_{\alpha\delta}g_{\beta\gamma}-g_{\alpha\gamma}g_{\beta\delta}, \\
R_{\mu\nu} &=& -2g_{\mu\nu}, \qquad  G_{\mu\nu} =  g_{\mu\nu}.
\end{eqnarray}
In 3 dimensions, the Riemann tensor further obeys the following relation
\begin{eqnarray}
\label{3dreim}
 R_{\alpha\beta\gamma\delta} &=& \epsilon_{\alpha\beta\rho}\epsilon_{\gamma\delta\sigma}( 
R^{\rho\sigma} -  \frac{1}{2}R g^{\rho\sigma} ), \\ \nonumber
&=& \epsilon_{\alpha\beta\rho}\epsilon_{\gamma\delta\sigma}G^{\rho\sigma} .
\end{eqnarray}
Here  $G^{\rho\sigma}$ is the Einstein tensor and the 
epsilon tensor is defined as
\begin{equation}
\epsilon^{\alpha\beta\gamma}=\frac{\widetilde{\epsilon}^{\ \alpha\beta\gamma}}{\sg}, \qquad \widetilde{\epsilon} ^{\ +\xi -} = 1.
\end{equation}
where, \( \widetilde{\epsilon}^{\alpha\beta\gamma}\) is the completely antisymmetric Levi-Civita symbol.  The  epsilon tensor in 3 dimensions satisfies the relation
\begin{equation}
\label{epident}
 \epsilon^{\ \ \, \alpha}_{\beta\rho} \epsilon_{\alpha\delta\sigma} = - 
( g_{\beta\delta}g_{\rho\sigma} - g_{\beta\sigma}g_{\rho\delta} ) .
\end{equation}
Now using the definition of the flat space gamma matrices given in (\ref{defgamma}) 
and the vierbein in (\ref{vier}) it is easy to see that the 
the curved space gamma matrices satisfy 
\begin{equation}\label{idengam}
 [ \Gamma^\mu, \Gamma^\nu] = 2 \epsilon^{\mu\nu\rho}\Gamma_\rho, \qquad
\Gamma^\mu\Gamma_\mu =3. 
\end{equation}

\subsection*{Relation between $\nabla^2$ and $\ns^2$ in $AdS_3$}

To establish the relation between the spin $s$ Laplacian  and $\ns^2$  we  start by 
 considering  the action of $\ns^2$ on the object $\Psi_{\mu_1\mu_2\dots\mu_n}$. 
\begin{eqnarray} \label{the-one}
\ns^2 \Psi_{\mu_1\mu_2\dots\mu_n} 
& =& \Gamma^\mu \Gamma^\nu \nabla_\mu \nabla_\nu \Psi_{\mu_1\mu_2\dots\mu_n},  \nn \\
&=& \frac{1}{2} \{\G^\mu,\G^\nu \}  \nabla_\mu \nabla_\nu \Psi_{\mu_1\mu_2\dots\mu_n}  + \frac{1}{2}[\G^\mu,\G^\nu]  \nabla_\mu \nabla_\nu \Psi_{\mu_1\mu_2\dots\mu_n},   \nn \\
&=& \nabla ^2 \Psi_{\mu_1\mu_2\dots\mu_n} + \frac{1}{4}[\G^\mu,\G^\nu]  [\nabla_\mu , \nabla_\nu ] \Psi_{\mu_1\mu_2\dots\mu_n}. 
\end{eqnarray}
Thus, we require to evaluate the second term in the  above expression.
\begin{align}  \label{term2}
  & \frac{1}{4}[\G^\mu,\G^\nu]  [\nabla_\mu , \nabla_\nu ] \Psi_{\mu_1\mu_2\dots\mu_n} \nn \\
=& \frac{1}{32} R_{\mu\nu\sigma\delta} [\G^\mu,\G^\nu]   [ \G^\sigma , \G^\delta ]  \Psi_{\mu_1\mu_2\dots\mu_n}  \nn \\
&\qquad +\frac{1}{4}   [\G^\mu,\G^\nu]  g^{\eta\alpha}  \big{(} R_{\alpha \mu_1\nu\mu} \Psi_{\eta\mu_2\dots\mu_n} + R^\eta_{\alpha \mu_2\nu\mu} \Psi_{\mu_1\eta\mu_3\dots\mu_n} + \cdots + R^\eta_{\alpha \mu_n\nu\mu} \Psi_{\mu_1\mu_2\dots\mu_{n-1}\eta}  \big{)}, \nn \\
=&  \frac{1}{32} \mathcal{G}_{\mu_1\mu_2\dots\mu_n} + \frac{1}{4}  \mathcal{H}_{\mu_1\mu_2\dots\mu_n} .
\end{align}
To obtain the first equality we have used the definition of the covariant derivative 
given in (\ref{defcova}). 
We then define
\begin{align}
 \mathcal{G}_{\mu_1\mu_2\dots\mu_n} &=  R_{\mu\nu\sigma\delta} [\G^\mu,\G^\nu]   [ \G^\sigma , \G^\delta ]  \Psi_{\mu_1\mu_2\dots\mu_n}, \\
  \mathcal{H}_{\mu_1\mu_2\dots\mu_n} &= [\G^\mu,\G^\nu]  g^{\eta\alpha}  \big{(} R_{\alpha \mu_1\nu\mu} \Psi_{\eta\mu_2\dots\mu_n} + R_{\alpha \mu_2\nu\mu} \Psi_{\mu_1\eta\mu_3\dots\mu_n} + \cdots \nn \\
  &\qquad \qquad\qquad \qquad+ 
R_{\alpha \mu_n\nu\mu} \Psi_{\mu_1\mu_2\dots\mu_{n-1}\eta}  \big{)}.
\end{align}
Let's evaluate \(  \mathcal{G}_{\mu_1\mu_2\dots\mu_n}  \) first.  
\begin{align}\label{gee}
\mathcal{G}_{\mu_1\mu_2\dots\mu_n} &=  R_{\mu\nu\sigma\delta} [\G^\mu,\G^\nu]   [ \G^\sigma , \G^\delta ]  \Psi_{\mu_1\mu_2\dots\mu_n}, \nn \\
&= (g_{\mu\delta} g_{\nu\sigma} -  g_{\mu\sigma} g_{\nu\delta} )( 2\epsilon^{\mu\nu\alpha} \G_\alpha )  ( 2\epsilon^{ \sigma\delta\beta }\G_\beta )  \Psi_{\mu_1\mu_2\dots\mu_n}, \nn \\
&= 8( \delta^\eta_\eta g^{\alpha\beta}  \G_\alpha \G_\beta  - \delta^\alpha_\eta g^{\eta\beta}  \G_\alpha \G_\beta) \Psi_{\mu_1\mu_2\dots\mu_n}, \nn \\
&= 8( 3\cdot 3 - 3  ) \Psi_{\mu_1\mu_2\dots\mu_n} 
= 48 \ \Psi_{\mu_1\mu_2\dots\mu_n} .
\end{align}
In performing these manipulations we have used  the indentities in (\ref{epident}) and (\ref{idengam}). 
Let's now evaluate  \(  \mathcal{H}_{\mu_1\mu_2\dots\mu_n}\). 
\begin{align}
&\mathcal{H}_{\mu_1\mu_2\dots\mu_n} \nn \\
=&  [\G^\mu,\G^\nu]  g^{\eta\alpha}  \big{(} R_{\alpha \mu_1\nu\mu} \Psi_{\eta\mu_2\dots\mu_n} + R_{\alpha \mu_2\nu\mu} \Psi_{\mu_1\eta\mu_3\dots\mu_n} + \cdots + R_{\alpha \mu_n\nu\mu} \Psi_{\mu_1\mu_2\dots\mu_{n-1}\eta}  \big{)}, \nn \\
=& g^{\alpha\eta} [\G^\mu,\G^\nu]  \sum_{j=1}^{n} (g_{\alpha\mu}g_{\mu_j\nu} - g_{\alpha\nu}g_{\mu_j\mu} ) \Psi_{\eta\mu_1\dots\check{\mu}_j\dots\mu_n}, \nn \\
=&  2\G^\mu\G^\nu  \sum_{j=1}^{n}   g_{\mu_j \nu} \Psi_{\mu\mu_1\dots\check{\mu}_j\dots\mu_n}, 
\label{eich}
= 2(2g^{\mu\nu} - \G^\nu\G^\mu )  \sum_{j=1}^{n}   g_{\mu_j \nu} \Psi_{\mu\mu_1\dots\check{\mu}_j\dots\mu_n}, \nn \\
=& 4g^{\mu\nu}  \sum_{j=1}^{n}   g_{\mu_j \nu} \Psi_{\mu\mu_1\dots\check{\mu}_j\dots\mu_n}  = 4  \sum_{j=1}^{n}  \Psi_{\mu_1\dots\mu_j\dots\mu_n}, \nn \\
=& 4 n  \Psi_{\mu_1\mu_2\dots\mu_n}.
\end{align}
Here we have used the tracelessness condition (\ref{tr}) several times to simplify the terms. 
Using (\ref{gee}) and (\ref{eich}) in (\ref{term2}) we have
\begin{align}
   \frac{1}{4}[\G^\mu,\G^\nu]  [\nabla_\mu , \nabla_\nu ] \Psi_{\mu_1\mu_2\dots\mu_n} &= \frac{48}{32}  \Psi_{\mu_1\mu_2\dots\mu_n} +   \frac{4n}{4}  \Psi_{\mu_1\mu_2\dots\mu_n}, \nn \\
   &= \left(n+\frac{3}{2}\right)  \Psi_{\mu_1\mu_2\dots\mu_n}, \nn \\
   &= \left(s+1 \right)  \Psi_{\mu_1\mu_2\dots\mu_n}. 
\end{align}
Thus from (\ref{the-one}) we have
\begin{equation}
\ns^2  \Psi_{\mu_1\mu_2\dots\mu_n}  = \left( \nabla^2 + (s+1) \right)  \Psi_{\mu_1\mu_2\dots\mu_n}.
\end{equation}

\section{Diagonalization of the mass matrix: Proof }\label{app-A}

In this section we will prove Identity 2 (\ref{claim-1}) which is given by 
\begin{align}
 T^{(n)}M^{(n)}[T^{(n)}]^{-1} &=  D^{(n)},  \\
M^{(n)}[T^{(n)}]^{-1}&=[T^{(n)}]^{-1}D^{(n)},  \nn \\
 M^{(n)}T^{(n)}&=T^{(n)}D^{(n)} .
\end{align}
where, $ D^{(n)}_{pq} =(n-2p+m)\delta_{pq}$.  To arrive at  the last line we have used 
Identity 1 (\ref{claim-0}). 
This is equivalent to showing 
\begin{equation} \label{piden}
\sum_{b=0}^n \sum_{c=0}^n M^{(n)}_{ab}T^{(n)}_{bc} x^c =\sum_{b=0}^n \sum_{c=0}^n   T^{(n)}_{ab} D^{(n)}_{bc} x^c.
\end{equation}
\def\sp{{(n,a)}}
Let's start with the LHS of (\ref{piden}). 
and define the variable 
\begin{equation} 
 z=\frac{x+1}{x-1}.
\end{equation}
Then we obtain 
\begin{align}
&\sum_{b=0}^n  M^{(n)}_{ab}\sum_{c=0}^s T^{(n)}_{bc} x^c \nn \\
=& \sum_{b=0}^n  M^{(n)}_{ab} (x+1)^b (x-1)^{n-b},  \nn \\
=& (x-1)^n  \sum_{b=0}^n  M^{(n)}_{ab} z^b,  \nn \\
=& (x-1)^n  \sum_{b=0}^n [  - a\delta_{a-1,b} - (n-a)\delta _{a+1,b} + m \delta _{a,b} ] z^b, \nn \\
=& (x-1)^n [ -az^{a-1}-(n-a)z^{a+1}+mz^a ],  \nn \\
=& (x-1)^{n-a-1} (x+1)^{a-1} [(m-n)x^2 + 2(2a-n)x -(m+n)]. 
\end{align}
We now need to evaluate the RHS. Before proceeding we shall list a few definitions and identities which we shall be using. 
\begin{align}
P_{(n,a)}(x) &=\sum_{b=0}^n T^{(n)}_{ab} x^b = (x+1)^{a}(x-1)^{n- a}, \\ \nn
x\frac{dP_{(n,a)}(x)}{dx} &= \sum_{b=0}^n b T^{(n)}_{ab} x^b. 
\end{align}
For the RHS we have, 
\begin{align}
&\sum_{b=0}^n T^{(n)}_{ab}\sum_{c=0}^n D^{(n)}_{bc} x^c , \nn \\
=&  \sum_{b=0}^n T^{(n)}_{ab}\sum_{c=0}^n [n-2b+m]\delta_{bc} x^c,  \nn \\
=&  \sum_{b=0}^n T^{(n)}_{ab}  [n-2b+m] x^b,  \nn \\
=&  (n+m) \sum_{b=0}^n  T^{(n)}_{ab} x^b -2  \sum_{b=0}^n b T^{(n)}_{ab} x^b,  \nn \\
=& (n+m) P_{(n,a)}(x) -2x\frac{dP_{(n,a)}(x)}{dx, } \nn \\
=& (n+m) (x+1)^a(x-1)^{n-a} \nn \\ &-2x[a(x+1)^{a-1}(x-1)^{n-a} + (n-a)(x+1)^{a}(x-1)^{n-a-1}],  \nn \\
=& (x-1)^{n-a-1} (x+1)^{a-1} [(m-n)x^2 + 2(2a-n)x -(m+n)]. 
\end{align}
This completes the proof of (\ref{claim-1}). 

\providecommand{\href}[2]{#2}\begingroup\raggedright\endgroup


\end{document}